\documentclass[aps,showpacs,preprintnumbers,nofootinbib]{revtex4}
\usepackage{amssymb}
\usepackage{amsmath}
\usepackage{bm}

\begin{document}

\title{On Weyl cosmology in five dimensions and the cosmological constant }
\author{Jos\'{e} Edgar Madriz Aguilar\thanks{%
E-mail address: jemadriz@fisica.ufpb.br}, and Carlos Romero \thanks{%
E-mail address: cromero@fisica.ufpb.br} }
\affiliation{$^{a}$Departamento de F\'{\i}sica, Universidade Federal da Para\'{\i}ba,
Caixa Postal 5008, 58059-970 \\
Jo\~{a}o Pessoa, PB, Brazil\\
E-mail: jemadriz@fisica.ufpb.br; cromero@fisica.ufpb.br}

\begin{abstract}
In this talk notes we expose the possibility to induce the cosmological constant from extra dimensions, in a geometrical framework where our four-dimensional Riemannian space-time is embedded into a five-dimensional Weyl integrable space. In particular following the approach of the induced matter theory, we show that when we go down from five to four dimensions, we may recover in the context of the general theory of relativity, the induced energy momentum tensor of the induced matter theory plus a cosmological constant term, that is determined by the presence of the Weyl scalar field on the bulk. 
\end{abstract}

\maketitle

\section{Introduction}

Due to the recent discovery
of cosmic acceleration there has been a renewed interest in the role the
cosmological constant could play to explain the new data. For instance,
there is strong evidence that the so-called dark energy might have a connection with the
cosmological constant. Moreover, the present most popular model of
cosmology, the Lambda-CDM model, tacitally assumes the existence of the
cosmological constant \cite{tegmark}. Particle physics
theorists have always argued in favour of the existence of the cosmological
constant as a consequence of the energy density of the vacuum \cite{Weinberg}%
. From the standpoint of cosmological theory it seems then desirable to have a
justification of the cosmological constant on theoretical grounds. This
quest has led some theoreticians to modify Einstein's gravitational theory,
these attempts going back to the works of Eddington and Schr\"{o}dinger \cite{Goenner}.

In this talk notes, based on our recent work \cite{pa}, we present a new approach to this old
question in the context of extra dimensions \cite{Arkani}, \cite{ndGR},\cite{Ponce}. In particular in our proposal
our ordinary spacetime is viewed as a hypersurface (\textit{the
brane}) embedded in a five-dimensional (5D) manifold (\textit{the bulk}) which is described by a Weyl integrable geometry. Mathematical theorems regulate these embeddings, in
particular, the Campbell-Magaard theorem \cite{campbell} and its extensions
specify the conditions under which the embeddings are possible \cite{Tavakol}. In the case when the Weyl field depends only on the extra dimension, the embedded spacetime is Riemannian and general relativity holds in 
the brane \cite{we1}, although the non-Riemannian character of the whole
bulk propagates into the brane.

\section{The formalism}

Let us consider a five-dimensional space $M^{5}$ endowed with a metric
tensor $^{(5)}g$ and an integrable Weyl scalar field $\phi $. In local
coordinates $\{y^{a}$$\}$ the five-dimensional line element can be written as  
$ dS^{2}=g_{ab}(y)\,dy^{a}dy^{b}$, where \ $g_{ab}$ are the components of $^{(5)}g.$\ The simplest action that can be constructed for a Weylian theory of gravity
in \ a five-dimensional vacuum is given by 
\begin{equation}
^{(5)}\!\mathcal{S}=\int d^{5}y\,\sqrt{\left\vert ^{(5)}g\right\vert }\left[
^{(5)}\!\mathcal{R}+\xi \phi ^{a}\,_{;a}\right] ,  \label{w5d2}
\end{equation}%
where $\xi $ is an arbitrary coupling constant, $\phi _{a}\equiv \phi _{,a}$
is the gauge vector associated to the Weyl field $\phi (y)$, $\left\vert
^{(5)}g\right\vert $ is the absolute value of \ the determinant of the
metric $^{(5)}g_{ab}$, $^{(5)}\!\mathcal{R}$ is the Weylian Ricci scalar.
The variation of the action (\ref{w5d2}) with respect to the tensor metric and  the Weyl scalar field yields 
\begin{eqnarray}
&& ^{(5)}\!\mathcal{G}_{ab}+\phi _{a;b}-(2\xi -1)\phi _{a}\phi _{b}+\xi
g_{ab}\phi _{c}\phi ^{c}=0,  \label{w3} \\
&& \phi ^{a}\,_{;a}+2\,\phi _{a}\phi ^{a}=0  \label{w4}
\end{eqnarray}
where $^{(5)}\!\mathcal{G}_{ab}$ denotes the Einstein tensor calculated with
the Weyl connection $^{(5)}\Gamma _{bc}^{a}=^{(5)}\{_{bc}^{\,a}\}-(1/2)[\phi
_{b}\delta _{c}^{a}+\phi _{c}\delta _{b}^{a}-g_{bc}\phi ^{a}]$ and $%
\{_{bc}^{\,a}\}$ are the Christoffel symbols of Riemannian geometry. The
equations (\ref{w3}) and \ (\ref{w4}) are the field equations of the
five-dimensional Weyl gravitational theory and describes the dynamics of a
five-dimensional bulk in vacuum. A better insight may be gained if we recast
the field equations (\ref{w3}) and (\ref{w4}) into its Riemannian part plus
the contribution of the Weyl scalar field \cite{Novello}. 
We are then led to 
\begin{eqnarray}
&&^{(5)}\tilde{G}_{ab}-\frac{1}{2}(6-5\xi )\left[ \phi _{a}\phi _{b}-\frac{1%
}{2}g_{ab}\phi _{c}\phi ^{c}\right] =0,  \label{nueva5D3} \\
&&^{(5)}\tilde{\Box}\phi =0,
\end{eqnarray}%
where the tilde $(\sim )$ is used to denote quantities calculated with the
Riemannian part of the Weyl connection and $^{(5)}\tilde{\Box}$ denotes the
5D d'Alembertian operator in the Riemannian sense. Now expressing the local coordinates $\{y^{a}\}$ as $ \{x^{\alpha },l\}$ and denoting  by $l$ the fifth (spacelike) coordinate, we
choose for simplicity a line element in the form \footnote{%
We shall adopt the convention $diag(+---)$ for the signature of $g_{\alpha
\beta }.$} 
\begin{equation}
dS^{2}=g_{\alpha \beta }(x,l)dx^{\alpha }dx^{\beta }-\Phi ^{2}(x,l)dl^{2}.
\label{metric}
\end{equation}%
where the function $\Phi ^{2}(x,l)$ is the 5D analogue of the lapse function
used in canonical general relativity, which supposes that spacetime may be
foliated by a family of spacelike surfaces \cite{Wheeler}. As in the induced matter approach \cite{Ponce}, with respect to the geometry given by (\ref{metric}) the field equations (\ref{nueva5D3}) can be splitted as
\begin{eqnarray}
&&^{(5)}\tilde{G}_{\alpha \beta }-\frac{1}{2}(6-5\xi )\left[ \phi _{\alpha
}\phi _{\beta }-\frac{1}{2}g_{\alpha \beta }\left( \phi _{\gamma }\phi
^{\gamma }-\Phi ^{-2}\phi _{l}^{2}\right) \right] =0,  \label{nueva5D6a}\\
&& ^{(5)}\tilde{G}_{\alpha l}-\frac{1}{2}(6-5\xi )\phi _{\alpha }\phi _{l}=0,
\label{nueva5D7a}\\
&& ^{(5)}\tilde{G}_{ll}-\frac{1}{4}(6-5\xi )\left[ \phi _{l}^{2}+\Phi ^{2}\phi
_{\gamma }\phi ^{\gamma }\right] =0.  \label{nueva5D8a}
\end{eqnarray}
Some solutions of the above field equations have been worked out in detail by
Novello and colaborators considering different geometric settings \cite%
{Novello}. 
Motivated by the recently result \cite{we1}, if the Weyl scalar field depends only on the extra coordinate $l$, then each leaf of the foliation $l=const$ (which represents our 4D spacetime) has a Riemannian character and can be locally and isometrically embedded in a five-dimensional Weylian space whose metrical properties are given by (\ref{metric}). 
In view of the above let us assume that $\phi =\phi (l)$, i.e\ the Weyl
scalar field depends only on the extra coordinate $l$. In this case the
field equations (\ref{nueva5D6a}), (\ref{nueva5D7a}) and (\ref{nueva5D8a})
become 
\begin{eqnarray}
&&^{(5)}\tilde{G}_{\alpha \beta }+\frac{1}{4}(5\xi -6)\Phi ^{-2}g_{\alpha
\beta }\phi _{l}^{2}=0,  \label{nueva5D7} \\
&&^{(5)}\tilde{G}_{\alpha l}=0,  \label{nueva5D8} \\
&&^{(5)}\tilde{G}_{ll}-\frac{1}{4}(6-5\xi )\phi _{l}^{2}=0,  \label{nueva5D9}
\\
&&\frac{\partial }{\partial l}\left[ \sqrt{\left\vert g_{5}\right\vert }%
\,\Phi ^{-2}\phi _{l}^{2}\right] =0.  \label{nueva5D10}
\end{eqnarray}%
On a particular hypersurface $\Sigma _{0}$ the induced
line element will be given by $
dS_{\Sigma _{0}}^{2}=h_{\alpha \beta }(x)dx^{\alpha }dx^{\beta }$, 
where $h_{\alpha \beta }(x)=g_{\alpha \beta }(x,l_{0})$ is the induced
metric on $\Sigma _{0}$. From the Gauss-Codazzi equations it is easy to show
(See, for instance \cite{Israelit}) that the induced dynamics on the
hypersurface $\Sigma _{0}$ is governed by the four-dimensional field
equations 
\begin{equation}
^{(4)}\tilde{G}_{\alpha \beta }=T_{\alpha \beta }^{(IM)}+\Lambda
(x)h_{\alpha \beta },  \label{nu4d1}
\end{equation}
where $T^{(IM)}_{\alpha \beta}$ is the energy momentum tensor
obtained in the induced matter approach \cite{Ponce}, which has the form  
$ T_{\alpha \beta }^{(IM)}=(\Phi _{\alpha ||\beta }/\Phi )+(1/
2\Phi ^{2})\{ (\overset{\star }{\Phi }/\Phi )\overset{\star }{g}
_{\alpha \beta }-\overset{\star \star }{g}_{\alpha \beta }+g^{\lambda \mu }
\overset{\star }{g}_{\alpha \lambda }\overset{\star }{g}_{\beta \mu }-(1
/2)g^{\mu \nu }\overset{\star }{g}_{\mu \nu }\overset{\star }{g}_{\alpha
\beta } + (1/4)g_{\alpha \beta }[ \overset{\star }{g}^{\mu \nu }
\overset{\star }{g}_{\mu \nu }+( g^{\mu \nu }\overset{\star }{g}_{\mu
\nu }) ^{2}] \}$ ,  
with the bars $(||)$ denoting covariant derivative in a Riemannian sense and
the star $(\star )$ denoting derivative with respect to the fifth coordinate 
$l$, and the function $\Lambda (x)$ is given by 
\begin{equation}
\Lambda (x)=\frac{1}{4}(6-5\xi )\Phi ^{-2}\left. \phi _{l}^{2}\right\vert
_{l=l_{0}}.  \label{nu4d3}
\end{equation}%
The induced energy-momentum tensor $T_{\alpha \beta
}^{(IM)}$\ can be obtained even if the bulk is Riemannian, but the
interesting fact here is that the function $\Lambda (x)$ is a new
contribution depending directly on the Weyl scalar field. It is worth
mentioning that when the lapse function $\Phi $ dependends only on the time
then $\Lambda (t)$ can be interpreted as an induced cosmological parameter,
whereas if $\Phi $ is constant then (\ref{nu4d3})  reduces to an induced
cosmological constant.\\
\section{An example}

As an application of the formalism let us show some interesting results. To follow the details in deriving this results the reader may check our work \cite{pa}.\\
When we consider a 5D space in the form of a triple
warped product manifold \cite{Dahia1} with metric given by 
\begin{equation}
dS^{2}=dt^{2}-a^{2}(t)dr^{2}-e^{2F(t)}dl^{2},  \label{nueva5D11}
\end{equation}
where $dr^{2}=\delta _{ij}dx^{i}dx^{j}$ is the three-dimensional Euclidian
line element, $t$ represents the cosmic time for co-moving observers, $F(t)$
is a well-behaved real function and $a(t)$ is the cosmological scale factor, the 4D induced line element on $\Sigma _0$ becomes a FRW one. The metric function $F(t)$ is determined by the system (\ref{nueva5D7}) to (\ref{nueva5D10}) and when a power law expanding universe $a(t)\sim t^{p}$ is regarded, it gives $F(t)=ln (Bt^{\gamma})$ where $B$ is a constant and $\gamma=(1/2-p)+(1/2)\sqrt{1-32p^{2}+16p}$. To have real values for the
power $\gamma $ that are compatible with an expanding universe ($p>0$), the
values of $p$ must range in the interval $0<p\leq (1/4)+\sqrt{6}/8$. For (\ref{nueva5D11}) the equation (\ref{nueva5D10}) gives a Weyl scalar field $\phi(l)=C_{1}l+C_{2}$, with $C_1$ and $C_2$ integration constants. Consequently using (\ref{nu4d3}) we can induce a variable cosmological ``constant'' $ \Lambda (t) =\left( \frac{C_{1}}{2}\right) ^{2}(6-5\xi
)B_{1}^{-2}t^{-2\gamma }$. Clearly for $\gamma >0$ we have a decaying cosmological constant which may be a good candidate to explain the present period of accelerated expansion of the universe.

\section{Final Remarks}

In this talk notes we have exposed the idea of generating a cosmological
constant, or rather, a cosmological parameter, from extra dimensions.
Although this has already been investigated in the context of induced matter
theory, the novelty of our approach is to regard the same problem in\ a more
general setting, i.e by assuming the geometry of the embedding space to have
a Weylian character. Two comments are in order: Firstly, the embedding space
has a prescribed dynamics; secondly, the embedding does not affect the
Riemannian geometry of the spacetime. These features depend on the fact that
the Weyl field is assumed to be integrable and depending only on the extra
dimension. Finally, by setting up a simple\ "toy model " our intention is to
call attention to the richness of non-Riemannian geometries, in particular
to the Weyl integrable manifolds, as a way of providing new degrees of
freedom that might play a role in the theoretical framework of
higher-dimensional embedding theories of spacetime. We believe that in this
context issues such as the nature of the cosmological constant, dark energy
and other important questions may be investigated from an entirely new point
of view.

\section*{Acknowledgements}

\noindent The authors would like to thank CNPq-CLAF and CNPq-FAPESQ (PRONEX)
for financial support. We are indebted to Dr. F. Dahia for helpful comments.

\end{document}